\documentstyle[12pt,aps,preprint]{revtex}
\newcommand{\be}{\begin{equation}}
\newcommand{\ee}{\end{equation}}
\newcommand{\ba}{\begin{array}{c}}
\newcommand{\ea}{\end{array}}
\newcommand{\bqa}{\begin{eqnarray}}
\newcommand{\eqa}{\end{eqnarray}}

\begin{document}
\draft
\title{Studying High Energy Final State Interactions by N/D Method}
\author{Feng Yuan and Hanqing Zheng}
\address{Department of Physics, Peking University, Beijing 100871,\\
People's Republic of China}
\maketitle

\begin{abstract}
We discuss the final state interaction effects at high energies via a
multi-channel N/D method. We find that the 2 by 2 charge--exchange final
state interactions typically contribute an enhancement factor of a few times 
$10^{-2}$ in the $B$ meson decay amplitudes, both for the real and the
imaginary part. We also make some discussions on  the elastic rescattering
effects.
\end{abstract}


\vspace{1cm}

There have been increasing interests in recent years in studying the final
state interaction (FSI) effects in B meson hadronic decays. In Ref.~\cite{hz}
, a Regge pole model was suggested to estimate the FSI effects in B decays
and it was found that, for $B\to DK$ decays the FSI effects are small in the
sense that the corresponding partial wave $S$--matrix is close to unity.
Especially the final state rescattering effects in charge-exchange processes
are of order $O(10^{-2})$. These results imply that for the color
non-suppressed charged final states, a tree level calculation to the B decay
amplitude can be a good approximation~\cite{nardulli}. Using a similar
method as in Ref.~\cite{hz}, the authors of Ref.~\cite{donoghue} observed
that, as a consequence of Pomeron exchanges the FSI effects do not vanish as
the center of mass energy square $s$ approaches infinity and therefore they
suggest that there are sizable strong phases generated from FSIs. On the
other hand, the authors of Ref.~\cite{blok} pointed out that in above
estimates the real part contributions of the rescattering amplitudes were
not considered. They use a dispersion relation to estimate also the real
part contributions of the charge--exchange processes ($B\to D\pi $,$\pi \pi $
) and find that Reggeon contributions (in charge--exchange rescattering
processes) are one order of magnitude larger than those estimated in Ref.~
\cite{hz,nardulli}. Since this result, if correct, leads to important
phenomenological consequences~\cite{rosner}, it is therefore worthwhile to
re-examine the problem by carefully studying the FSI effects, including the
real part, of charge--exchange rescattering processes, which is what this
paper mainly devote to. We will also discuss the FSI effects of elastic
rescatterings although there exist some uncertainties related to such
rescatterings in the scheme presented in this paper.

To correctly evaluate the FSI effects to the decay amplitudes, it is
important to note that if the initial interactions are of short range the
production amplitude and FSIs are actually factorized~\cite{watson}.
Assuming (quasi) two--particle unitarity we will use the multi-channel N/D
method~\cite{bjorken} to study the FSI effects. The inelasticity effects
caused by those multi--particle states is presumably not large, since the
pion multiplicity is rather low at $s=m_B^2$ as the multiplicity increases
only as $\log (s)$ (see \cite{blok} and references therein) and at least
part of the multi--particle final states can be classified as two--particle
final states of resonances (accompanied by cascade resonance decays).

For a multi-channel system we can rewrite the partial wave T-matrix as, 
\begin{equation}
{\bf {T=N/D}\ ,}  \label{nd}
\end{equation}
where the bold face represents a matrix and ${\bf {N}}$ contains only the
left--hand singularities and ${\bf {D}}$ the right--hand singularities. In
the physical region, ${\bf {T}}$ and the production amplitude, ${\bf {A}}$,
satisfy the following unitarity relations, 
\begin{equation}
{\rm Im}{\bf {T}={T}^{*}{\rho }{T},\,\,{\rm Im}{A}={A}^{*}{\rho }{T}\ ,}
\end{equation}
which hold when the center of mass energy square, $s$ takes any value along
the positive real axis above the lightest threshold in the complex $s$
plane, if there is no anomalous threshold. We will come back to discuss the
influence of the anomalous thresholds on high energy FSIs later in this
paper and will argue that their effects are negligible. The unitarity
relations lead to the following simultaneous integral equations for ${\bf {N}
}$ and ${\bf {D}}$, 
\begin{equation}
{\bf N}={\frac 1\pi }\int_L{\frac{{\bf b}(s^{\prime }){\bf D}(s^{\prime })}{
s^{\prime }-s}}ds^{\prime }\ ,  \label{NDI} \\
{\bf D}={\bf 1}-{\frac 1\pi }\int_R{\frac{{\bf \rho }(s^{\prime }){\bf N}
(s^{\prime })}{s^{\prime }-s}}ds^{\prime }\ ,
\end{equation}
where ${\bf {b}}$ is the discontinuity of ${\bf {T}}$ from the left--hand
cut. Equation (\ref{NDI}) may be solved by iteration method.

The FSI contribution to the production amplitude obeys a simple formula, 
\begin{equation}
{\bf A}={\bf A}_0/{\bf D}\ ,  \label{fd}
\end{equation}
where ${\bf A}$ is the full decay amplitude and ${\bf A}_0$ is analytic in a
region surrounding the positive real $s$ axis. In the present physical
situation, it implies ${\bf A}_0$ is a smooth function of $s$ in a large
domain surrounding $s=m_B^2$ in the presence of inelasticity. It should be
noticed that the dispersion integrals in Eq.(\ref{NDI}) for those components
of {\bf N} and {\bf D} in which Pomeron exchanges involve should receive one
subtraction. In potential scattering theory {\bf D} is normalized to unity
when $s$ equals to infinity. Since the total cross sections of
hadron--hadron scatterings are continuously rising as a function of $s$, it
is impossible to have such a normalization condition for high energy hadron
interactions. ${\bf A}_0$ in Eq.~(\ref{fd}) is the Born term in potential
scattering theory and be constant. In the present case there is an
uncertainty in relating ${\bf A}_0$ to the bare amplitude. This is related
to the multiplicative ambiguity in N/D method and the lacking of a natural
normalization point in {\bf D}. However, as will be seen below, this
ambiguity can actually be avoided in calculating charge--exchange
rescatterings.

To extract the information on charge--exchange processes from Eq.~(\ref{fd})
we may rewrite {\bf D} as, 
\begin{equation}
{\bf D}={\bf \overline{D}}-{\bf D^c}\ ,
\end{equation}
where ${\bf \overline{D}}$ contains only channels dominated by Pomeron
exchanges and ${\bf D^c}$ represents charge--exchange processes. Assuming $
{\bf D^c}$ to be a small quantity and that ${\bf \overline{D}}$ does not
deviate far from unity (these assumptions will be justified by the results
given later), we can make an expansion in the denominator of Eq.~(\ref{fd})
in powers of ${\bf D^c}$ and keep only the leading term to obtain, 
\begin{equation}
{\bf A}_{charge-exchange}={\bf A}_{el}{\bf D^c}/{\bf \overline{D}}\ ,
\label{fdc}
\end{equation}
where ${\bf A}_{el}$ is the physical amplitude in the limit of vanishing
charge--exchange rescatterings: 
\begin{equation}
{\bf A}_{el}={\bf A}_0/{\bf \overline{D}}\ ,
\end{equation}
which can be a good approximation to the physical amplitudes for decay
channels with large(larger) bare amplitude. Equation (\ref{fdc}) can be
written in an explicit form after a little algebra, 
\begin{equation}
{\bf A}_{ch.-ex.}={\bf A}_{el}\{{\frac 1\pi }\int {\frac{{\bf \rho }
(s^{\prime }){\bf T}^c(s^{\prime }){\bf {\cal D}}^P(s^{\prime })}{s^{\prime
}-s}}ds^{\prime }+{\frac 1\pi }\int {\frac{{\bf \rho }(s^{\prime }){\bf T}
^P(s^{\prime }){\bf {\cal D}}^c(s^{\prime })}{s^{\prime }-s}}ds^{\prime }\}\
,  \label{equ}
\end{equation}
where $s=m_B^2$ and, 
\begin{equation}
{\bf {\cal D}}^P(s^{\prime })={\frac{\overline{{\bf D}}(s^{\prime })}{
\overline{{\bf D}}(s)}}\ ,\,\,\,\,{\bf {\cal D}}^c(s^{\prime })={\frac{{\bf D
}^c(s^{\prime })}{\overline{{\bf D}}(s)}}\ .  \label{df}
\end{equation}
It is straightforward to check that ${\bf {\cal D}}^P$ satisfies the
following integral equation, 
\begin{equation}
{\bf {\cal D}}^P(s^{\prime })=1-{\frac{s^{\prime }-s}\pi }\int {\frac{{\bf 
\rho {\ T}}^P(s^{\prime \prime }){\bf {\cal D}}^P(s^{\prime \prime
})ds^{\prime \prime }}{(s^{\prime \prime }-s^{\prime })(s^{\prime \prime }-s)
}}\ ,  \label{ds}
\end{equation}
and Eqs.~(\ref{fdc}), (\ref{equ}) and (\ref{df}) define an integral equation
for ${\bf {\cal D}}^c$, provided ${\bf T}^P$, ${\bf T}^c$ and ${\bf {\cal D}}
^P$ are known. The above equations are valid to the leading order in the
expansion in powers of ${\bf T}^c$ and to all orders of Pomeron exchanges.

In order to calculate the FSI contributions in charge--exchange processes it
is required to determine ${\bf {\cal D}}^P(s^{\prime})$ through Eq.~(\ref{ds}
). It is a difficult task since for a process $i\to j$ under consideration
there can be many channels ($j^{\prime}\to j$) dominated by Pomeron exchange
exist contributing to $i\to j$ through $i \to j^{\prime}\to j$. To proceed
let's look at the elastic process ($j^{\prime}=j$) first and approximate the
solution of Eq.~(\ref{ds}) by first order iteration result, one obtains, 
\begin{equation}  \label{app}
{\cal D}_{el}(s^{\prime})=1- {\frac{s^{\prime}-s}{\pi}}\int {\frac{\rho
T^P_{el}(s^{\prime\prime})ds^{\prime\prime}}{(s^{\prime\prime}-s^{
\prime})(s^{\prime\prime}-s)}}\ ,
\end{equation}
where 
\begin{equation}
T^P_{el}(s)={\frac{i\beta^P}{16\pi s_0(\lambda +\alpha^{\prime}\log(s/s0)
-i\pi\alpha^{\prime}/2)}}\ ,
\end{equation}
where $\beta^P$ is the Pomeron coupling to matter and $\lambda =2.82$ is the
parameter characterizing the form-factor of the Pomeron coupling, also we
set $\alpha_0^P=1$ in above. From Eq.(\ref{app}) we find that ${\cal D}^P(s)$
is a smooth function of $s$ and deviates very little from unity for any
reasonable value of $s$. Even for $s$ being as large as $W$ boson mass
square, one still has, 
\begin{equation}
{\cal D}_{el}(M_W^2) = 0.99+6.1\times 10^{-2}i
\end{equation}
in $D\pi$ case. The {\bf D} matrix elements of inelastic diffractive
channels are expected to be the same order as $D_{el}-1\sim 10^{-2}$ but
with different phases. Large cancellation is natural when summing over all
the intermediate states $j^{\prime}$ which ensures terms like $%
T^c_{ij^{\prime}}T^P_{j^{\prime}j}$ in Eq. (\ref{equ}) remains to be second
order comparing with $T^c_{ij}$. Therefore we can approximate Eq.~(\ref{equ}
) by, 
\begin{equation}  \label{equ2}
{\bf A}_{ch.-ex.}={\bf A}_{el} \{ {\frac{{\rm P}}{\pi}} \int {\frac{{\bf \rho
}(s^{\prime}){\bf T}^c(s^{\prime})}{s^{\prime}-s}}ds^{\prime}+i{\bf \rho}
{\bf T}^c \}\ ,
\end{equation}
which is, in simple words, obtained by neglecting all the diffractive
scattering effects in charge--exchange rescatterings.\footnote{
This equation was used in Ref.\cite{prd} without any justification.}
Equation (\ref{equ2}) is the basic formula for numerical calculations in the
following.\footnote{
For $\pi\pi$ rescatterings, ${\cal D}_{el}(M_W^2)=0.95+0.20i$, which is not
as close to unity as in the $D\pi$ case. This is not surprising as the
diffractive FSI effects are stronger for $\pi\pi$ rescatterings. However,
Eq.~(\ref{equ2}) should still be good enough to work with in an order of
magnitude estimate.}

In order to compare with previous work, we now focus on the process, $B\to
D^0\pi ^0$ through a $D^{+}\pi ^{-}$ intermediate state. The $D^{+}\pi
^{-}\to D^0\pi ^0$ scattering amplitude is, 
\begin{equation}
T_{D^{+}\pi ^{-}\to D^0\pi ^0}=2\sqrt{2}\rho \ ,  \label{1}
\end{equation}
where the $\rho $--Reggeon contribution is parameterized as~\cite{hz}, 
\begin{equation}
\rho =\beta ^R(t)\left( {\frac s{s_0}}\right) ^{\alpha ^R(t)}{\frac{
1-e^{-i\pi \alpha ^R(t)}}{\sin \pi \alpha _0^R}}\ ,  \label{2}
\end{equation}
in the small negative $t$ region.
We take $s_0\simeq 2\alpha ^{\prime }{}^{-1}$ and also neglect the $t$
dependence of $\beta ^R$, according to Ref.~\cite{blok}. The Reggeon
coupling adopted in Ref.~\cite{blok} is obtained by relating $\beta _R$ to
the on-shell vector meson coupling in the $t$--channel physical process
which however over-estimates the magnitude comparing with the one directly
extracted from high energy scattering data. We take the value used in Ref.~
\cite{blok} for numerical calculation and will come back to this difference
later.

The approximation made above will be self-consistent if the final results on
FSI effects are small which is indeed the case as we find, 
\begin{equation}  \label{res}
{\frac{A(B\to D^+\pi^-\to D^0\pi^0)}{A(B\to D^+\pi^-)}}= (-6.9+4.4i)\times
10^{-2}\ .
\end{equation}
Similarly for the $\pi^+\pi^-\to \pi^0\pi^0$ rescattering, we have, 
\begin{equation}  \label{11}
T_{\pi^+\pi^-\to \pi^0\pi^0}=4\rho \ ,
\end{equation}
in which $s_0=\alpha^{\prime -1}$ is taken and the numerical result is, 
\begin{equation}  \label{res2}
{\frac{A(B\to \pi^+\pi^-\to \pi^0\pi^0)}{A(B\to \pi^+\pi^-)}}\simeq
(-10.+5.5i)\times 10^{-2}\ .
\end{equation}
These results are significantly smaller in magnitude than the one obtained
in Ref.~\cite{blok}.
Still, a more careful fit to the high energy scattering data suggest that
the $\rho$-Reggeon coupling is further suppressed by a factor about 2.5~\cite
{nardulli} comparing with the previously used. With this suppression factor
the $\beta_R$ parameter is close to the value used in Ref.~\cite{hz}
\footnote{
When extracting $\beta_R$ from high energy scattering data it is probably
better not to include in the fit the nucleon--nucleon scattering data which
contain strong SED violating contributions and can not be explained by a
simple Regge pole model. Actually, the fact that meson--nucleon data looks
better in favor of a simple Regge pole parameterization is the main reason
to support the use of a simple Regge pole model for meson--meson
scatterings. On the other hand, absorption effects, if exist, only reduce
the low partial wave FSI effects.}. We therefore suggest that Reggeon
contributions to FSIs are typically of order of a few times 10$^{-2}$, both
for the real parts and the imaginary parts.

Of course Regge parameterization can not always be correct when decreasing $
s $ down to thresholds even though we know from old experiences that it may
indeed work well in charge--exchange processes at rather low energies. To
estimate the violation of the Regge parameterization which we call $\delta$,
one has, 
\begin{equation}
\delta\simeq {\frac{\bar s {\overline {\rho T^c}}}{\pi m_B^2}}< {\frac{\bar s
}{2 \pi m_B^2}}\ ,
\end{equation}
where ${\overline{\rho T^c}}$ refers to the averaged value of the magnitude
of $\rho T^c$ in the range of $\bar s$ which characterizes the region where
Regge parameterization fails and the last inequality comes from unitarity.
Taking $\bar s=10GeV^2$ we see that $\delta $ is at best a few percent, in a
very conservative estimate.

Strictly speaking, the present method only applies to a restricted class of
T matrices which contain no anomalous threshold. An anomalous threshold
comes from dynamical singularities which occurs only if a loosely bound
composite system of hadronic constituents is involved~\cite{oehme}. The
typical mass scale of an anomalous threshold is the scale of the masses of
the light hadrons in the decay products~\cite{mandelstan}. From a simple
dimensional analysis we are convinced that these effects, if appear, are
further suppressed in powers of $1/s$ in high energy scatterings. 

In above we have discussed the effects of charge--exchange FSIs, we obtained
the relation between the rescattering amplitude and the physical amplitude
in the limit of vanishing charge--exchange rescatterings (the bare amplitude
renormalized by diffractive rescatterings). The ambiguity caused by the
normalization of {\bf D} is absorbed into $A_{el}$. In discussing the
elastic rescattering it is unavoidable to face such an ambiguity. As having
been made clear in above discussions that the ambiguity comes from the fact
that there is no natural normalization point for {\bf D}, unlike the case in
potential scattering theory where {\bf D} is normalized to 1 at $s=\infty $.
The only thing can be done in the present formalism is to compare the FSI
strength between two energy scales.
Here we define the FSI strength at $s$ ($=m_B^2$) to be $1/{\bf D}_\Lambda
(s)$ where ${\bf D}_\Lambda $ is normalized to unity at scale $\Lambda $,
and $\Lambda $ is chosen somewhere below W boson mass and much larger than $
m_B$.\footnote{
See the discussion given in Ref.~\cite{suzuki}.} From the previous
estimation on ${\bf {\cal D}}_{el}(M_W^2)$, recall that $D(s_1)$ normalized
at $s_2$ is equal to the inverse of $D(s_2)$ normalized at $s_1$, we
conclude that the FSI effects remain small even for elastic rescatterings at 
$s=m_B^2$, at least in the $D\pi $ case.\footnote{
Of course, a more careful numerical study calls for a calculation beyond the
first order iteration in determining function ${\cal {\bf D}}$, including
inelastic effects which can only be done under various assumptions and
approximations. The random S--matrix approach~\cite{suzuki} can be helpful
in extracting out the essential feature of a system with many inelastic
channels.} Here we defined in some way (maybe rather artificial) the
absolute value of the phase of the elastic channel, but we want to stress
that what truly matters in a physical quantity (like the CP violation
observables) is the difference between two elastic phases of different
isospins which is of non-diffractive nature and should therefore be a small
quantity.

To conclude we find that charge--exchange rescatterings in B decays
contribute typically an enhancement factor of order of a few times $
10^{-2}\sim \lambda^2$ where $\lambda $ is the Wolfenstein parameter. For
elastic rescatterings, the FSI effects most likely remain small. B decays
contain rich phenomenology and an enhancement factor $\sim \lambda$ via
charge--exchange rescatterings leads an important role~\cite{rosner}. In
this sense the result of this paper, if correct, is in disfavor of an
experimental extraction to the rich physical phenomena.

{\it Acknowledgment}: It is our pleasure to thank Valeri Markushin and
Hong-An Peng for helpful discussions.
The work of H.Z. is supported in part by China
National Natural Science Foundations.

\end{document}